\documentclass[12pt]{iopart}
\usepackage{graphicx}

\begin{document}
\title[Born-Infeld
Phantom Gravastars]{
Born-Infeld
Phantom Gravastars
 }
 \author{Neven Bili\'{c}\dag
\footnote[3]{Author to
whom correspondence should be addressed},
Gary B Tupper\ddag\ and Raoul D Viollier\ddag}
 \address{\dag\
 Rudjer Bo\v skovi\'c Institute,
P.O. Box 180, 10002 Zagreb, Croatia}
\address{\ddag\
Institute of Theoretical Physics and Astrophysics,
 Department of Physics, University of Cape Town,
 Private Bag, Rondebosch 7701, South Africa}

 \ead{
 \mailto{bilic@thphys.irb.hr},
 \mailto{parsons@physci.uct.ac.za},
 \mailto{viollier@physci.uct.ac.za}}
%
\begin{abstract}
We construct new gravitational vacuum star solutions with a Born-Infeld phantom replacing the de Sitter interior. The model allows for a wide range of masses and radii required by phenomenology, and can be motivated from low-energy string theory.
\end{abstract}
%
\maketitle

\section{Introduction}
From its inception, the Schwarzschild solution of Einstein's equations
has been the subject of controversy over the possibility of
compressing matter within the gravitational radius $2GM$.
Given the accumulated evidence for supermassive compact objects  ranging from
a few $10^{6}M_{\odot}$ to a few $10^{9} \, M_{\odot}$,
the existence of black-hole-like objects is beyond doubt \cite{melia,korm}
 What does
remain an issue is whether the Schwarzschild metric correctly describes
the physics of the interior.
Alternatives to classical black holes have been proposed
with no singularities in the interior
\cite{viol5,viol6,chap1,chap2,mazu3,dym,mazu4,math,lob3,ben,bro}.

The simplest model proposed  for supermassive compact objects at the
galactic centers is a self-gravitating
degenerate fermion gas composed of, e.g.,
 heavy sterile neutrinos
 \cite{viol5,viol6,bil7,bil8,mun9}
   However, this scenario
cannot cover the whole mass range of supermassive black-hole
candidates with a single sterile neutrino
species \cite{bil1}.

Recently, Chapline et al \cite{chap1,chap2}  put forth an
interesting proposal based on analogies to condensed matter
systems where the effective general relativity was an emergent phenomenon.
Specifically, they suggested that the sphere
 where the lapse function vanished marked a quantum phase
transition, the lapse function increasing again at $r < 2GM$.
As this required negative pressure, the authors of \cite{chap2}
assumed the interior vacuum condensate to be described by de Sitter space
with the equation of state $p=-\rho$.

Subsequently, the idea of gravitational vacuum condensate, or `gravastar',
 was taken up by Mazur and Mottola \cite{mazu3,mazu4},
replacing the horizon with a shell of stiff matter astride
the surface at $r=2GM$.
Visser and Wiltshire \cite{viss4} and recently Carter \cite{car} also examined the stability
of the gravastar using the Israel thin shell formalism \cite{isra5}.
Despite the fact that general relativity is an emergent phenomenon
in string theory \cite{gree6}, the gravastar has met with a cool reception.
Certainly, the assumption of a de Sitter interior presents a quandary:
on the one hand, the quantum phase transition would suggest that
the associated cosmological constant is a fundamental parameter;
on the other hand, to accommodate the mass range of supermassive black hole
candidates, it must vary over some six orders of magnitude.
Thus, it seems prudent to explore other possibilities for the gravastar interior.


One obvious extension of the de Sitter equation of state
$p+\rho=0$ would be an equation  that satisfies
$p+\rho \leq 0$,
  thus violating the dominant energy condition.
The fluid of which the equation of state violates the dominant energy condition
has been
dubbed {\em phantom energy} \cite{cald1,cald2}
and has recently become a popular alternative
to quintessence and to the
cosmological constant \cite{phantom}.
The motivation for introducing the phantom energy is that
the equation of state $w\equiv p/\rho < -1$ produces a superaccelerated
cosmological expansion  which seems to be favored by the combined analysis of
the CMB and super nova type I data \cite{super}.
However,  a superaccelerated expansion may also be obtained
without violating the dominant energy condition
in scalar-tensor theories of gravity \cite{polar}
and in models with variable gravitational constant \cite{stef}
or variable cosmological constant \cite{sola}.   
Some  astrophysical aspects of the phantom have recently been discussed,
such as phantom energy accretion \cite{bab},
phantom energy wormholes \cite{lob2},
 and
a possible relation of the phantom tachyon model to
supermassive black holes \cite{gonz}.

In this paper we consider a gravastar interior consisting of a self-gravitating scalar field described by a  Born-Infeld type Lagrangian which 
 yields the Chaplygin gas equation of state \cite{kamen7,bil2}.
Hence, we look for static solutions of the self-gravitating Chaplygin gas.
In particular, we consider static Chaplygin gas configurations in the phantom regime, 
i.e., when $p+\rho<0$, and we show that these configurations could
provide an alternative scenario for compact massive objects at galactic
centers.

The paper is organized as follows: In section \ref{model} we introduce the
basics of the model. In section \ref{static} we investigate static 
solutions using Tolman-Oppenheimer-Volkoff equations.
In section \ref{galactic} we discuss a possible interpretation of
galactic centers as Born-Infeld phantom gravastars.
A stability analysis is given in section \ref{stability} and we conclude the paper
by section \ref{conclude}

\section{The model}
\label{model}
Consider the equation of state
\begin{equation}
p = - \; \frac{A}{\rho}
\label{eq006}
\end{equation}
in the phantom regime, i.e., when 
\begin{equation}
\rho < \sqrt{A} \, .
\label{eq106}
\end{equation}
 Equation (\ref{eq006}) describes the Chaplygin gas which,
for $\rho \geq \sqrt{A}$, has attracted some attention as a
dark energy candidate \cite{kamen7,bil2}.
Astrophysical objects made of the so-called {\em generalized Chaplygin gas} \cite{bento}
have recently been discussed \cite{bert}.
The generalized Chaplygin gas has also been exploited in the phantom
regime \cite{hao2}.
As we shall shortly demonstrate, static solutions
to Einstein's equations with matter described by (\ref{eq006}) with (\ref{eq106})
cover the range of masses and radii required to fit
the phenomenology of supermassive dark compact objects
at the galactic centers. Moreover, Eq. (\ref{eq006})
yields the de Sitter gravastar solution
in the limit when the central density of the static solution
approaches the value $\sqrt{A}$.

The Chaplygin gas equation of state (\ref{eq006}) with the condition
$\rho < \sqrt{A}$ may be derived from
the Dirac-Born-Infeld type Lagrangian
\begin{equation}
{\cal{L}}_{\rm DBI} = - \sqrt{A}\, \sqrt{ 1 + X}\, ,
\label{eq015}
\end{equation}
where
\begin{equation}
X= g^{\mu \nu}  \vartheta_{,\mu}
 \vartheta_{,\nu} \: .
\label{eq115}
\end{equation}
Clearly,  in the limit $X\rightarrow0$
this Lagrangian
becomes a free scalar field Lagrangian with a ``wrong sign"
kinetic term, hence a phantom\footnote{It has recently been shown that quantum effects could lead to an effective
 dark-energy equation of state violating the dominant energy condition on
cosmological scales even  for a scalar field Lagrangian having 
the correct sign of the kinetic energy \cite{onemli}.}.
A  phantom Lagrangian of the type (\ref{eq015}) has been proposed in the
context of the
superaccelerated expansion \cite{hao}.
Note that the Lagrangian (\ref{eq015})
leads to a perfect fluid with  the four-velocity
\begin{eqnarray}
u_{\alpha} = \frac{\vartheta_{, \alpha}}{\sqrt{ X}} \: ,
\label{eq016}
\end{eqnarray}
the pressure $p={\cal{L}}_{\rm DBI}$, and
 the density
\begin{eqnarray}
\rho = \frac{\sqrt{A}}{\sqrt{ 1 + X}}
 \leq  \sqrt{A}
 \: .
\label{eq017}
\end{eqnarray}
Clearly, the pressure and the density
obey Eq. (\ref{eq006}) and $\rho\leq  \sqrt{A}$.
It is worth noting that
the Chaplygin gas cosmological model \cite{kamen7}, in which
the condition
$\rho>\sqrt{A}$ holds, is described by
the  Lagrangian (\ref{eq015})
with a ($-$) sign in front of $X$.
It is as though instead of the lapse function
changing its sign, as it does in the Schwarzschild case for $r < 2 G M$,
the scalar kinetic energy changed its sign to become a phantom.

Equivalent to ${\cal{L}}_{\rm DBI}$ of Eq. (\ref{eq015}) is
\begin{equation}
{\cal{L}} = - \frac{\phi^{2}}{2} \, X -
\frac{1}{2} \; \left( \phi^{2} + \frac{A}{\phi^{2}} \right) ,
\label{eq019}
\end{equation}
as may be seen by eliminating $\phi^{2}$ through its equation of motion.
Now recall that in four dimensions, a three-form field strength is dual to a scalar:
\begin{equation}
H^{\mu \nu \alpha} = \phi^{2} \; \epsilon^{\mu \nu \alpha \beta} \; \vartheta_{, \beta} \; \; .
\label{eq020}
\end{equation}
Then the content of the model is
\begin{equation}
{\cal{L}} =  \frac{1}{12 \phi^{2}} \; H_{\mu \nu \alpha} \;
H^{\mu \nu \alpha} -
\frac{1}{2} \; \left( \phi^{2} + \frac{A}{\phi^{2}}  \right) \; \; .
\label{eq021}
\end{equation}
Notably, this is also the content of low-energy string theory
\cite{gree6} in the Einstein frame where
$H_{\mu \nu \alpha}$ is the Kalb-Ramond field, $\phi^{2}$
corresponds to the string coupling and the
dilaton kinetic term is neglected in lieu of a potential
assumed to arise from non-perturbative effects\footnote{Strictly,
 the exterior should have 
 $H_{\mu \nu \alpha} \; H^{\mu \nu \alpha} = 6A$ and $\phi^{2} = 0$ to be
Schwarzschild.}.
A potential similar to that in (\ref{eq019}) and (\ref{eq021})
has been considered in the context of polymer scaling and black holes
\cite{rama}.

\section{Static solutions}
\label{static}
Next, we proceed to solve Einstein's equations for a
static, spherically symmetric configuration with an interior
described by (\ref{eq006}) and with a Schwarzschild exterior.
Our approach is similar to that of Armendariz-Picon and Lim \cite{arm}.
However, in contrast to us, they consider
the classes of Lagrangian with strictly space-like gradient of $\phi$,
i.e. with $X<0$, so that in their case the quantity (\ref{eq016})  cannot be
interpreted as a four velocity. As a consequence, their energy-momentum tensor
does not have a standard perfect fluid form and the pressure is not isotropic.
Here, we consider strictly time-like gradient of $\phi$ with $X>0$, but we 
allow  phantom-like Lagrangians which violate the dominant energy condition.
 
For the static, spherically symmetric line element
\begin{equation}
ds^{2} = \xi^{2}(r) \; dt^{2} - \frac{dr^{2}}{ 1 - 2G{\cal{M}}(r)/r} - r^{2} \;
\left( d\vartheta^2 + \sin^{2} \vartheta \; d \varphi^{2} \right)
\label{eq002}
\end{equation}
with
$T_{\mu}^{\nu} = \mbox{diag} \;
( \rho  ,  -p  , - p  , - p)$,
 Einstein's equations become
\cite{oppen1}
\begin{equation}
{\cal{M}}' = 4 \pi r^{2} \; \rho ,
\label{eq003}
\end{equation}
\begin{equation}
\xi' = G \; \xi \;
\frac{ {\cal{M}} + 4 \pi \; r^{3} \; p}{r (r - 2 G \; {\cal{M}})} \, ,
\label{eq004}
\end{equation}
while $T_{\mu \; ; \nu}^{\nu} = 0$ gives
\begin{equation}
p' = - ( \rho + p ) \; \frac{\xi'}{\xi} \, .
\label{eq005}
\end{equation}
We focus on the equation of state
(\ref{eq006})
to close the system 
and  we require the solution to (\ref{eq003}) and (\ref{eq004}) to be regular
 at $r=0$.
 Since by rescaling $t$
one may set $\xi (0) = 1$, Eqs. (\ref{eq006}) and (\ref{eq005}) yield
\begin{equation}
\xi (r) = \frac{\rho}{\rho_{0}} \,
\sqrt{ \frac{A - \rho_{0}^{2}}{A - \rho^{2}} } \, .
\label{eq007}
\end{equation}
Combining Eqs.
(\ref{eq006}),
(\ref{eq004}),
and (\ref{eq005}),
 one has
\begin{equation}
\rho' = G \left( 1 - \frac{\rho^{2}}{A} \right)
\left( \frac{ \rho {\cal{M}} - 4 \pi A r^{3} }{r (r - 2G {\cal{M}} ) }  \right) \, .
\label{eq008}
\end{equation}

\begin{figure}
\begin{center}
\includegraphics[width=.5\textwidth,trim= 0 2cm 0 2cm]{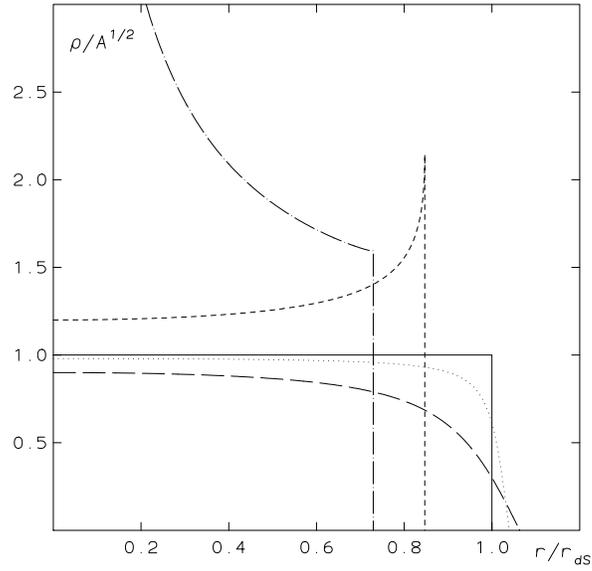}
\caption{
 Density profile of the Chaplygin star for
 $\rho_0/\sqrt{A}= 1.2$ (short dashed), 0.98 (dotted),
 0.9 (long dashed).
The limiting singular solution with
 $r^{-2/3}$ behavior at small $r$ is represented by the dot-dashed
 and the de Sitter gravastar by the solid line.
}
\label{fig1}
\end{center}
\end{figure}

\begin{figure}
\begin{center}
\includegraphics[width=.5\textwidth,trim= 0 2cm 0 2cm]{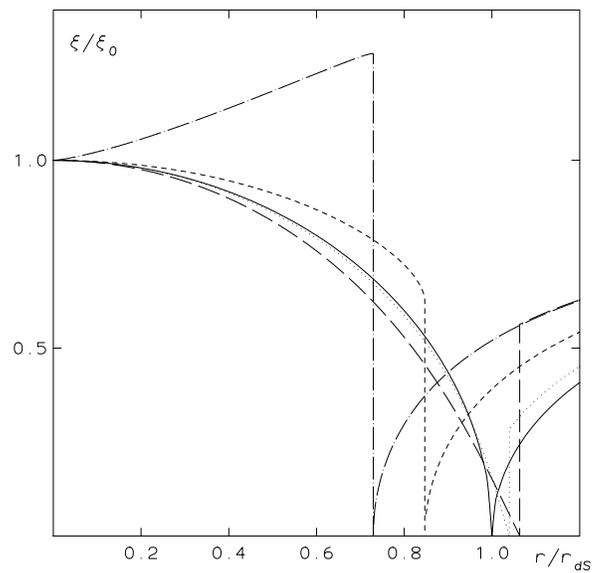}
\caption{
Lapse function $\xi/\xi_0$ for various solutions  as in Fig.\ \ref{fig1}.
}
\label{fig2}
\end{center}
\end{figure}

\begin{figure}
\begin{center}
\includegraphics[width=.5\textwidth,trim= 0 2cm 0 2cm]{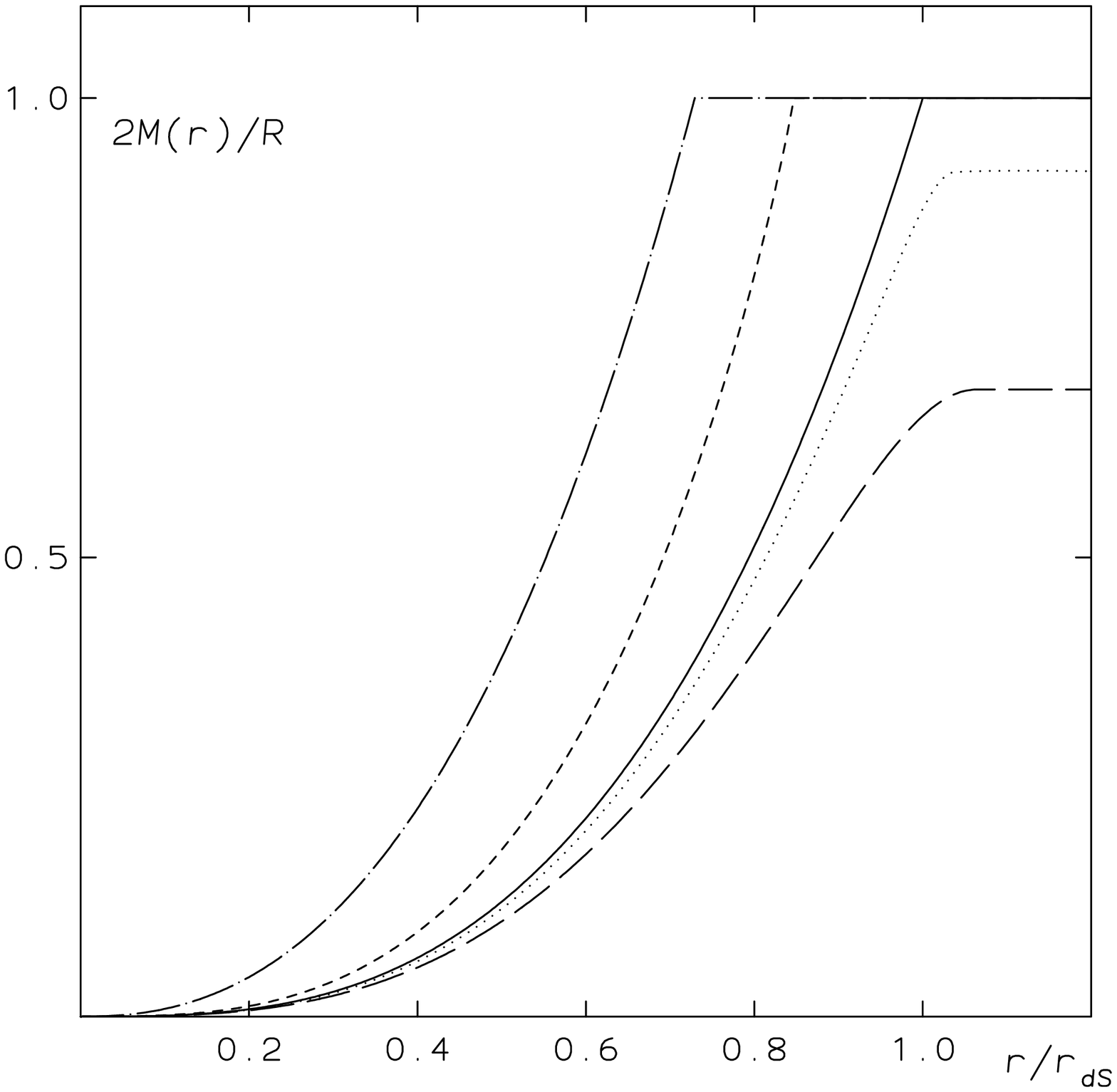}
\caption{
Enclosed mass divided by the radius of the star for
various solutions as in Fig.\ \ref{fig1}.
}
\label{fig3}
\end{center}
\end{figure}

In Figs. \ref{fig1}, \ref{fig2}, and \ref{fig3}, respectively, we exhibit
the resulting
$\rho (r) / \rho_{0}$, $\xi (r)$, and $2 G {\cal{M}} (r) / r$ for
selected values of
$\rho_{0} / \sqrt{A}$.
The solutions depend essentially
on the magnitude of $\rho_0$ relative to
 $\sqrt{A}$. In the following we summarize the properties of three classes of
 solutions corresponding to whether $\rho_0$ is larger, smaller, or
 equal to $\sqrt{A}$.

{\bf i)} For $\rho_0>\sqrt{A}$, the density
$\rho$ increases and the lapse function $\xi$ decreases with $r$
starting  from the origin
up to the black-hole horizon radius
 $R_{\rm bh}$,
    where $2G{\cal{M}}(R_{\rm bh})=R_{\rm bh}$.
In the limit
$\rho_0 \rightarrow \infty$,
a limiting
solution exists with a singular behavior
\begin{equation}
    \rho(r) \simeq (\frac{7 A}{18\pi G r^2})^{1/3}
\label{eq100}
\end{equation}
  near the origin.

{\bf ii)} For $\rho_0<\sqrt{A}$,
both
  $\rho$ and $\xi$ decrease with $r$ up to the radius $R_0$ where they vanish.
  At that point the pressure $p$ blows up to $-\infty$ 
     owing to (\ref{eq006}).
    The enclosed mass
${\cal M}$ is always less than $r/(2G)$,  never reaching the black-hole horizon,
i.e., the radius where $2G{\cal M}(r)=r$.

{\bf iii)} For $\rho_0=\sqrt{A}$,
the density $\rho$ remains
 constant equal to $\sqrt{A}$ up to  the de Sitter radius
  $R_{\rm dS}=2GM=8 \pi G R_{\rm dS}^3\rho_0/3$.
Hence, the interior is de Sitter, precisely as in Chapline et al.\ \cite{chap2}.
The lapse function is given by
\begin{equation}
\xi=\left(1-\frac{r^2}{R_{\rm dS}^2}\right)^{1/2}
\label{eq101}
\end{equation}
with
\begin{equation}
    R_{\rm dS} = \sqrt{\frac{3}{8\pi G}} A^{-1/4}.
\label{eq102}
\end{equation}

As $\rho_0 \rightarrow \sqrt{A} $  from above or from below,
solutions {\bf i)} or
{\bf ii)}, respectively, converge   to
{\bf iii)}
except at the endpoint.
The lapse function in {\bf iii)}
 joins the Schwarzschild solution outside
\begin{equation}
\xi(r)=\left(1-\frac{2 GM}{r}\right)^{1/2},
\label{eq103}
\end{equation}
continuously,
whereas in {\bf i)} and {\bf ii)} it happens discontinuously.

As in the case of a de Sitter gravastar, in order to 
join
our interior solutions to a
Schwarzschild exterior at a spherical boundary of radius $R$, 
 it is necessary to put a thin spherical shell\footnote{It has  recently
 been demonstrated that the joining can be made continuous
without the presence of a thin shell 
for a gravastar made of the fluid with an 
anisotropic pressure \cite{cat}.}
at the boundary with a surface density 
and a surface tension
satisfying Israel's junction conditions \cite{isra5}.
In 
all the solutions discussed above the pressure 
is isotropic  and does not vanish
at the boundary.
Hence, the pressure at the boundary
must be  compensated by a negative surface tension of the membrane.
We postpone this issue for section \ref{stability} where we discuss
the stability of the solution.

\section{Black holes at galactic centers}
\label{galactic}
Case {\bf ii)}, together with {\bf iii)}, is of particular
interest as we would like to interpret the supermassive
compact 
dark
objects at the galactic centers in terms of
phantom energy rather than in terms of a classical black hole.
It is natural to
assume that the most massive such object is described by
the de Sitter gravastar, i.e., solution {\bf iii)}
(depicted by the solid line in Figs. 1, 2, and  3).
If we identify the most massive black-hole candidate
observed at
the center of M87, with mass
$M_{\rm max}=
3\times 10^9 M_{\odot}$, with the de Sitter gravastar,
then $A^{1/8}=9.7 {\rm keV}^4$, to be contrasted with
the $10^{-3} {\rm eV}$ values wanted for  cosmology
\cite{bil2}.
The radius of this object is $R_{\rm dS}$ equal to the Schwarzschild radius
 $2GM_{\rm max}$.
Clearly,  solutions belonging to class {\bf ii)}, can fit
all masses $M<M_{\rm max}$.
However, for the phenomenology  of supermassive
galactic centers
it is important to find, at least approximately,
the mass-radius relationship for these solutions.
This may be done in the low central density approximation, i.e.,
$\rho_0 \ll \sqrt{A}$,
which is similar to the Newtonian approximation but,
in contrast to the  Newtonian approximation, one cannot neglect the pressure  term
in Eq. (\ref{eq004}). Moreover,
 as may be easily shown, in this approximation
${\cal{M}} \ll r^3 p$, so that the pressure term becomes dominant.
Next, neglecting $2G \cal M$ with respect to $r$, as in the usual
Newtonian approximation,
 Eq. (\ref{eq008})
simplifies to
$
\rho' =
 4 \pi G A r $,
\label{eq201}
with the solution
\begin{equation}
\rho = \rho_0 \left(
1-\frac{r^2}{R_0^2}\right) \, ;
\;\;\;\;\;
R_0^2= \frac{\rho_0}{2 \pi G A} \, ,
\label{eq202}
\end{equation}
which gives a mass-radius relation
\begin{equation}
\frac{M}{R_0^5}= \frac{16\pi^2}{15}GA={\rm constant}.
\label{eq205}
\end{equation}

The mass-radius relationship
$M \propto R_0^{5}$ which phantom gravastars obey, offers the prospect
of unifying the description of all supermassive compact dark objects that have been
observed at the galactic centers, as Born-Infeld phantom gravastars
with masses ranging from
$M_{\rm min} = 10^{6}  M_{\odot}$ to
$M_{\rm max} = 3 \times 10^{9}  M_{\odot}$.
Indeed, assuming that the most massive compact dark object, observed at
the center of M87, is a Born-Infeld phantom gravastar near the 
black-hole limit, with
$R_{\rm max} = 2GM_{\rm max} = 8.86 \times 10^{9} \mbox{km} = 8.21 \; \mbox{lhr}$,
the compact dark object at the center of our Galaxy, with mass
$M_{\rm GC} =  3 \times 10^{6} M_{\odot}$, would have a radius $R_{\rm GC}$ = 2.06 lhr
if the scaling law (\ref{eq205}) holds. This radius is well below the distances
of the closest approach to Sgr A$^{*}$ which the stars SO-2
($R_{\rm min}$ = 17 lhr = 123 AU, \cite{eck}) and SO-16 ($R_{\rm min}$ = 8.32 lhr = 60 AU \cite{ghez})
recently had and
beyond which the Keplerian nature of the gravitational potential of Sgr A$^{*}$
is well established.

\section{Dynamical stability}
\label{stability}
None of the solutions discussed in
the preceding section 
will be stable unless there is a membrane,
e.g. in the form of a thin shell,
placed at the boundary with surface density
$\sigma$ and  surface tension $\theta$
satisfying Israel's junction conditions.
Following Visser and Wiltshire \cite{viss4}
we consider a dynamical thin shell 
allowed to move radially 
at the boundary of the phantom gravastar
  and discuss
under which conditions will a gravastar configuration will be stable.
A dynamical thin shell connecting
two general static
spherically symmetric spacetimes
have also  recently been considered \cite{lob1} in a slightly
different context. 

Israel's junction conditions
read \cite{isra5}
\begin{equation}
\left[ \left[ K_{a}^{b} - \delta_{c}^{b}  K_{c}^{c} \right] \right]
= 8 \pi G  S_{a}^{b}\, ,
\label{eq009}
\end{equation}
where $S_{a}^{b}$ = diag ($\sigma,  \theta,  \theta$) is the
surface stress energy,
and $[[f]]$ denotes the discontinuity in $f$ across the boundary,
i.e.,
\begin{equation}
\left[ \left[ f (r) \right] \right] =
\lim_{\epsilon \rightarrow 0}  \,
\left( f (R + \epsilon) - f (R - \epsilon) \right)\, .
\end{equation}  
 The tensor 
  $K_{ab}$ is the extrinsic curvature 
defined by 
\begin{equation}
K_{ab}=h_{a}^{c}h_{b}^{d} n_{d;c}
\label{eq109}
\end{equation}  
where $n_a$ is a spacelike unit vector orthogonal to the timelike boundary and $h_{ab}$ is
the induced metric on the shell.

The angular components of the extrinsic curvature  
may be easily calculated from (\ref{eq109})  yielding 
\begin{equation}
K_{\vartheta}^{\vartheta} = K_{\varphi}^{\varphi} = \frac{1}{r} \,
(\Delta+\dot R)^{1/2}  \, ,
\label{eq210}
\end{equation}
precisely as in \cite{viss4},
where the dot denotes the derivative with respect to the proper time
and
where
\begin{equation}
\Delta=
 1 - \frac{2 G {\cal{M}}}{r} \, . 
\label{eq110}
\end{equation}
The calculation of the time-time component $K_t^t$ is
slightly more involved and is most easily done
following Israel \cite{isra5}.
By making use of the Gauss normal coordinates and orthogonality
from (\ref{eq109}) it follows 
\begin{equation}
K_{\tau}^{\tau} =  u^au^b K_{ab} = -n_a u^b {u^a}_{;b}
\label{eq211}
\end{equation}
where $u^a$ is the four-velocity 
of a point on the shell. Its non-vanishing components
 for the metric
(\ref{eq002}) are
\begin{equation}
 u^t = \frac{1}{\xi}\left(
 1+\frac{\dot{R}}{\Delta}
  \right)^{1/2}; 
  \;\;\;\;\;
  u^r= \dot{R}\, .
\label{eq212}
\end{equation}
Similarly, the non-vanishing components of $n_a$ are
\begin{equation}
n_t = \frac{\xi \dot{R}}{\Delta^{1/2}}\, ;
    \;\;\;\;\;
  n_r=\frac{(\Delta+ \dot{R}^2)^{1/2}}{\Delta}\, .
\label{eq213}
\end{equation}
Using these expressions it may be shown that the four-acceleration 
$u^b {u^a}_{;b}$
satisfies \cite{isra5}
\begin{equation}
n_a u^b {u^a}_{;b}=
 -\frac{1}{(\Delta+ \dot{R}^2)^{1/2}}
(\ddot{R} +\Gamma^r_{ab} u^a u^b) . 
\label{eq214}
\end{equation}
A straightforward calculation yields
\begin{equation}
K_{\tau}^{\tau} = 
\frac{1}{(\Delta+ \dot{R}^2)^{1/2}}
\left[ \ddot{R}+\frac{\xi'}{\xi} (\Delta+ \dot{R}^2 )
+\left(\frac{\cal{M}}{r}\right)'\frac{\dot{R}^2}{\Delta}
\right].
\label{eq216}
\end{equation}

Then, from (\ref{eq009}) using (\ref{eq003}) and (\ref{eq004}) we find
\begin{equation}
4\pi\sigma =-\frac{1}{G}\left[\left[ \frac{\sqrt{\Delta+\dot{R}^2}}{R}
\right]\right] ,
 \label{eq301} 
\end{equation}
\begin{equation}
4\pi (\sigma-2\theta)= 
\left[\left[\frac{R^2\ddot{R}
+G{\cal{M}}-4\pi G R^3\rho
}{ G R^2\sqrt{\Delta+\dot{R}^2}}
+\frac{4\pi  R\sqrt{\Delta+\dot{R}^2}}{\Delta}\,(\rho+p)
\right]\right] .
\label{eq302} 
\end{equation}
To derive the stability condition a la Visser and Wiltshire \cite{viss4},
it is now sufficient to replace $\dot{R}^2$ 
by $-2 V(R)$ and $\ddot{R}$ by $-V'(R)$ in equations (\ref{eq301}) and
(\ref{eq302}), where $V(R)$ is a potential. 
The shell will be stable against small radial perturbation
if  there exists an equilibrium position $\bar{R}$ such that
 \begin{equation}
V(\bar{R})=0;
\;\;\;\;
V'(\bar{R})=0;
\;\;\;\;
V''(\bar{R})>0 .
\label{eq303} 
\end{equation}
Then, by choosing a suitable potential $V(R)$, equations (\ref{eq301}) and 
(\ref{eq302}) define a parametric 
 equation of state $\theta=\theta(\sigma)$ for the shell.
 
For the static shell and the metric (\ref{eq002}) we find

\begin{equation}
\left( 1 - \frac{2 G M}{R} \right)^{1/2} -
\left( 1 - \frac{2 G {\cal{M}} (R)}{R}\right)^{1/2}
= - 4 \pi G  R  \sigma ,
\label{eq013}
\end{equation}
and with the help of Eq. (\ref{eq004}) we obtain 
\begin{eqnarray}
M
\left(1 - \frac{2 G M}{R }\right)^{-1/2}
-
 \left({\cal{M}} (R)
-
\frac{4 \pi R^{3} A}{ \rho (R)}\right) \left( 1 -
\frac{2 G {\cal{M}}(R)}{R}\right)^{-1/2}
& &     \nonumber \\
=
4 \pi R^{2} ( \sigma - 2 \theta) ,
& & 
\label{eq014}
\end{eqnarray}
where $M$ is the total mass.
Note that if the joining is affected
at the point $R_0$
where
$\xi = 0$ and $p$ is infinite
(the point of naked singularity \cite{cat}), the surface density $\sigma$
is finite, whereas the surface tension 
$\theta \rightarrow - \infty$. Hence, to avoid the singularity
the shell must be placed at some $R < R_0$.

For the Newtonian gravastars discussed in the preceding section,
equations (\ref{eq301}) and 
(\ref{eq302}) may be considerably simplified. 
Using (\ref{eq202}),
 the approximation
$2G{\cal{M}}/R \ll 1$, and assuming that the boundary radius $R$ is close to
  $R_0$, i.e.,
\begin{equation}
y\equiv \frac{R_0-R}{R_0} \ll 1.
\label{eq00}
\end{equation}
 we find
\begin{equation}
\sigma=2\pi G A R_0^3 \frac{y^2}{\sqrt{1-2V}}
  \, ,
\label{eq05}  
\end{equation}
\begin{equation}
\theta = -\frac{1}{8\pi G R_0}\frac{\sqrt{1-2V}}{y}
-4\pi G A R_0^3\frac{yV}{\sqrt{1-2V}}
+\pi G A R_0^4\frac{y^2 V'}{1-2V}
  \, .
\label{eq06}  
\end{equation}
In general, the potential $V$ is  a function of both $R_0$ and $y$,
depending on the chosen equation of state $\theta=\theta(\sigma)$.
The dynamical stability is achieved if for any $R_0$, $V$ has a minimum
at some point $\bar{y}=1-\bar{R}/R$ which should not depend strongly on $R_0$
and at which $V(\bar{y})=0$.

Instead of postulating $\theta=\theta(\sigma)$ we can 
choose a desirable potential and determine the equation of state
{\em a posteriori}.
Static stability (indifferent balance) is obtained by setting $V\equiv 0$.
Depending on the potential $V(R_0,y)$,  this may still be a  complicated parametric
equation of state. However,  
for the static shell, i.e. $V\equiv 0$, the equation of state simplifies to
\begin{equation}
\theta =-\left(\frac{A R_0}{32\pi G\sigma}\right)^{1/2} .
\label{eq11}  
\end{equation}
This equation  describes a 2-dimensional generalized anti-Chaplygin gas
\cite{gor}
with the equation of state of the form
$p=-\theta= A'/\sigma^{\alpha}$, with $\alpha=1/2$. 


\section{Conclusions}
\label{conclude}
In conclusion, we have shown that replacing the de Sitter interior of the gravastar by a Born-Infeld phantom allows a wide range of gravastar mass and radii related to the central density, or equivalently to the velocity of the phantom scalar. We have demonstrated that if the constant $A$, as the only free parameter of the model, is fixed by assuming
that the most massive galactic center object
is the maximal (de Sitter) gravastar, then the model is able to explain all
 supermassive compact dark objects at the center of the galaxies.
Furthermore, as demonstrated above, the phantom gravastar model can lay claim to a connection with low-energy string theory.

\subsection*{Acknowledgment}
 The work of NB   was supported by
 the Ministry of Science and Technology of the
 Republic of Croatia under Contract
 No. 0098002 and partially supported through the
 Agreement between the Astrophysical sector, SISSA, and the
 Particle Physics and Cosmology Group, RBI.
Two of us (RDV and GBT) acknowledge grants from the South
African National Research Foundation (NRF GUN-2053794),
the Research Committee of the University of Cape Town,
and the Foundation for Fundamental Research (FFR PHY-99-01241).

\section*{References}


\end{document}